\documentclass{article}
\usepackage{LaThuileFPSpro}
\begin{document}
\title{ 
  Status and Perspectives of Dark Matter Searches
  }
\author{
  Jodi A. Cooley        \\
  {\em Stanford University} \\
  {\em Varian Physics Laboratory MS 4060} \\
  {\em 382 via Pueblo Mall} \\
  {\em Stanford, CA  94305-4060}
  }
\maketitle

\baselineskip=11.6pt

\begin{abstract}
In the last year many collaborations searching for the dark matter constituent have published results from their experiments.  Here I give a review of direct detection searches reported by the DAMA, KIMS, CDMS-II, EDELWEISS-I, CRESST-II and ZEPLIN-I collaborations.  I also outline the future plans of each collaboration as well as the XENON10 collaboration. 
\end{abstract}
\newpage
\section{Introduction}
The first evidence for the existence of dark matter came from the study of galaxy clusters by Zwicky\cite{zwicky} in 1933.  He observed that the motion of galaxies within the clusters was not consistent with the amount of matter.  The idea that unseen dark matter is responsible for large, peculiar galaxy velocities as well as other phenomena is now generally accepted.   Today most research in the field is focused on the nature of dark matter.   The most stringent constraints on model parameters have been achieved by combining recent WMAP results with other experiments\cite{WMAP}.  These constraints indicate that about a quarter of the energy density of the universe consists of nonbaryonic dark matter.

In order to accommodate the relic densities of dark matter needed to explain the observed universe, the dark matter constituents must have an interaction cross-section consistent with the weak interaction.  These weakly interacting massive particles (WIMPs) are one class of dark matter particles that could be accommodated in big bang cosmology\cite{jungman}.  Many models of new physics outside the Standard Model offer a candidate for the dark matter constitutent.  In supersymmetry this candidate is the neutralino ($\chi$)\cite{neutralino},  a linear combination of binos (\~{B}), winos (\~{W}$_{3}$) and higgsinos (\~{H}$_{1}^{0}$ and \~{H}$_{2}^{0}$).   Neutralinos scatter elastically with atomic nuclei at a rate given by
\begin{equation}
  Rate \sim Nn_{\chi}<\sigma_{\chi}>,
  \label{eqrate} 
\end{equation}
where N is the number of targets in the detector, $n_{\chi}$ is the local neutralino density, and $<\sigma_{\chi}>$ is the WIMP-nucleon scattering cross-section.

There are two different techniques for  detecting dark matter.   One technique, known as {\em indirect detection}, is to look for products of neutralino annihilations in the sun or earth.  Experiments such as AMANDA\cite{amanda}, Super-K\cite{superk} and EGRET\cite{egret} rely on this technique.  The other technique is called {\em direct detection}.    Experimenters using this technique look for evidence of neutralinos interacting with nuclei in their detector's target medium.  This paper will review experiments using the latter technique.
\section{Detection Strategies}
The current experiments employ one of two strategies for detecting dark matter.  The first strategy is to look for an annual modulation of WIMPs due to the earth's rotation around the sun as the sun travels through the solar system.   The difficulty with this method is that one looks for a small, time-varying signal on top of a huge background.  It requires a large detector volume and a stable, nonfluctuating background.  The advantage is one does not need to know the characteristics of the background as long as it is not time-varying.  The other strategy is to substantially reduce the background to near zero and detect WIMP-nucleon interactions.   The difficulty with this approach is characterizing the backgrounds and achieving an extremely low-background experimental environment. 

Globally  there are a variety of detector mediums used including crystals such as Ge, Si, CaWO$^{4}$ and liquid noble gasses such as Xe and Ar.  There are three observables from the interactions of WIMPs and the target medium:  ionization, phonons resulting from the interaction of the WIMP with a nucleon in the crystal lattice, and scintillation.
\section{Backgrounds}
Dark matter experiments are faced with a variety of background sources including products of cosmic ray interactions and radioactivity from the environment and detector materials.  Although the specifics of dealing with these backgrounds vary from one experiment to another,  the general techniques used are the same.

One common background are neutrons produced by cosmic ray muon interactions.  In an effort to reduce the number of these neutrons seen by detectors, dark matter experiments are sited in deep underground laboratories.   For example, the neutron rate at 15 meters water equivalent (mwe)
is approximately 1/(kg s).  However,  at 2000 mwe that rate reduces to approximately 1/(kg day).  Another technique employed by dark matter experiments to reject these neutrons is to surround the detectors with active muon vetos.

Neutrons from natural radioactivity are another background that dark matter experiments frequently encounter.  These neutrons result from ($\alpha$, n) reactions and spontaneous fission of uranium and thorium nuclides which are natural in all geological formations.  An ($\alpha$, n) reaction occurs when uranium or thorium decays giving off an $\alpha$ particle which can react with elements such as Al or Na producing a neutron.  The number and energy of neutrons produced from these reactions varies depending on the type of rock.   A passive shielding of hydrocarbons is typically used to reduce these backgrounds.

The final background I will discuss are photons which can come from environmental radioactivity or  radiocontaminants in the detector itself or its shielding.  To reduce this type of background experiments frequently use passive shielding such as ancient lead and copper.  In an effort to further reduce the photon background, the detectors and shielding are build using low-radioactivity materials.
\section{DAMA}
The DAMA experiment is located in the Gran Sasso National Laboratory, Italy at 3500 mwe.   It is currently the only direct detection dark matter experiment to claim observation of a signal.  The experiment consists of nine 9.7 kg NaI crystals which are each viewed by two photomulitplier tubes.  DAMA does not distinguish between a WIMP signal  and background events directly.  Rather, it looks for a signal from the amplitude of the annual modulation of WIMPs due to the earth's rotation around the sun as the sun travels though the WIMP halo of our galaxy.

\begin{figure}[t]
  \vspace{9.0cm}
  \includegraphics{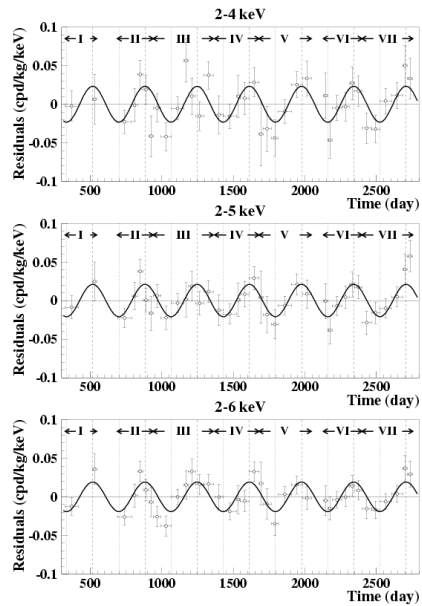}
  \caption{\it
    DAMA annual modulation signal\cite{dama}.  The fitted modulation of the data is shown in three energy intervals for seven years of data taking.   The claimed signal is at 6.3 $\sigma$.
    \label{damaFig} }
\end{figure}

DAMA's analysis technique\cite{dama} is to calculate the residual in each energy bin for single-hit events.  The residual is calculated by 
\begin{equation}
  R = <r_{ijk} - m_{jk}>
  \label{eqresidual} 
\end{equation}
where $r_{ijk}$ is the i-th time interval in the j-th detector for the k-th energy bin and $m_{jk}$ is the average over seven years in the j-th detector for the k-th energy bin.  Based on a fit to the sinusodial behaviour of the 2-6 keV energy bin, DAMA claims to have observed a positive WIMP signal at 6.3 $\sigma$ from seven years (107,731 kg days) of  observation.  Their best fit to the data is shown in Figure \ref{damaFig}.

The DAMA collaboration has been operating LIBRA, a 250 kg NaI experiment, since March 2003.  No results from this larger experiment had been released at the time of writing this document.
\section{KIMS}
The KIMS (Korea Invisible Mass Search) is located in the Yang Yang Underground Laboratory at a depth of 2000 mwe.  Their detector is similar to DAMA but uses CsI crystals, instead of NaI.  The most problematic internal background for KIMs was from $^{137}$Cs which was reduced by using purfied water during the crystal  production.  Their most recent result is based on one crystal with a mass of 6.6 kg and 237 kg days of data\cite{kims}.  As shown in figure \ref{limitsFig}, the KIMS experimental result just begins to explore the DAMA claimed signal region for spin-independent WIMP models.

KIMS is also working towards a larger detector.  Currently the detector is running four crystals with a mass of 8.66 kg each.  Two of the crystals have backgrounds of approximately six counts/keV/day and the remaining two have backgrounds of approximately four counts/keV/day.    Three more crystals are waiting for installation\cite{kimsFuture}.
\section{CDMS}
CDMS (Cryogenic Dark Matter Search) currently has the best limits on dark matter mass and cross-section in the field.   It is located at the Soudan Underground Laboratory in Minnesota, USA at a depth of 2090 mwe.  The CDMS detectors are made of Ge and Si crystals.  In contrast to the KIMS and DAMA experiments which use photomulitplier tubes to collect light, the CDMS detector crystals were photolithographically patterned with Al fins and W transition edge sensors (TESs) to collect athermal phonons and  electrodes to collect ionization charge.  

CDMS discriminates between background from $\gamma$-rays, electrons, and $\alpha$-particles which produce electron recoils, and WIMP and neutron events which produce nuclear recoils, using their ionization yield.  Ionization yield is defined as the ratio of ionization energy to recoil energy.  Electron events have yields around unity while nuclear events have yields around 0.3.  

\begin{figure}[!t]
  \vspace{9.0cm}
  \includegraphics{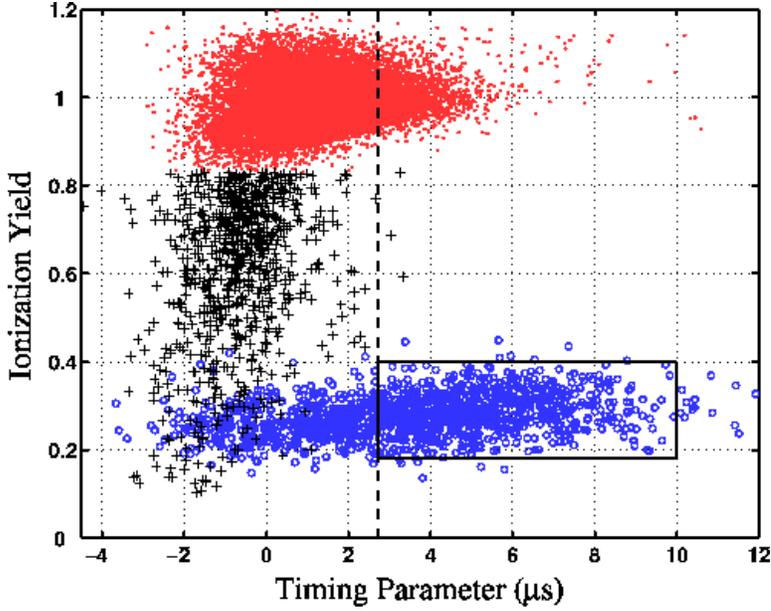}
  \caption{\it
    Ionization yield versus timing parameter for calibration data in Ge detector T2Z3 for events with recoil energies between 1 - 100 keV \cite{cdms}.  Typical bulk electron events from a $^{133}Ba$ calibration source are shown as red dots, low yield electron events are shown as black +, and neutron events from a $^{252}Cf$ source are shown as blue circles.   The dashed line vertical line indicates the minimum timing parameter allowed for WIMP candidates.   The box indicates the approximate signal region.
    \label{timingFig} }
\end{figure}

One class of background events that cannot be distinguished this way are electrons that interact near the surface of the detector.  These events suffer from incomplete charge collection resulting in a reduced yield.  Fortunately, these events produce a different frequency spectrum of phonons.  These phonons travel faster, resulting in  a shorter rise-time of the phonon pulse.   Hence, a minimun requirement on rise-time eliminates most of this troubling background.  Figure \ref{timingFig} demonstrates the rejection capability of such a cut.

To distinguish between neutron and WIMP events, the CDMS experiment uses two targets, Ge and Si.  Neutrons in the energy range of 50 keV to 10 MeV interact in silicon at twice the rate per kilogram compared to germanium.  By contrast, WIMPs have a six-times lower interaction rate in silicon compared to germanium. 

\begin{figure}[t]
  \vspace{9.0cm}
  \includegraphics{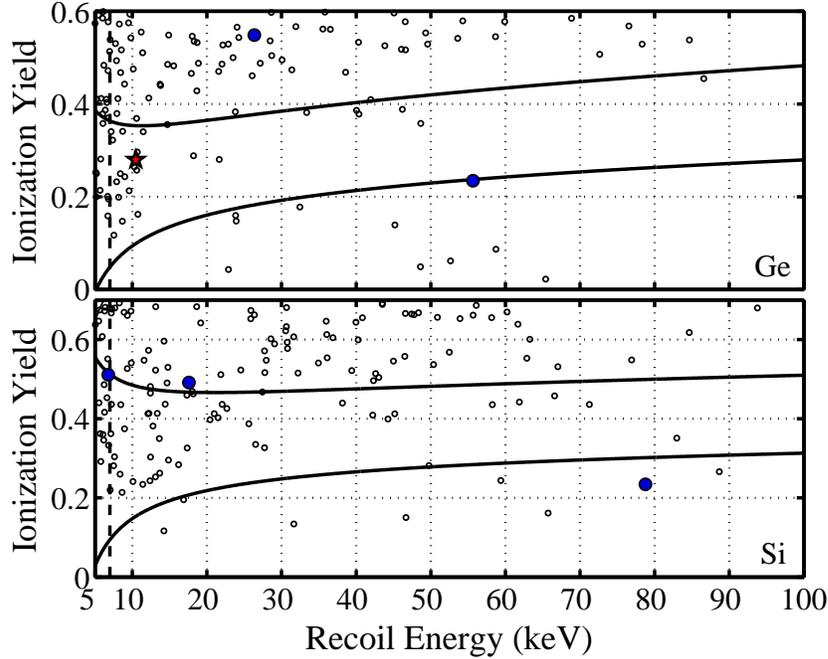}
  \caption{\it
  Ionization yield (ionization/recoil energy) vs recoil energy for all events in Ge (upper) and Si (lower) detectors passing initial data selection criteria.  The solid curved lines indicate the nuclear recoil band defined by a $^{252}Cf$ calibration source.The vertical dashed line indicates the 7 keV analysis threshold.  Events passing the surface electron timing cut outside the nuclear recoil band are indicated by blue dots.  The event that passed the surface electron timing cut inside the nuclear recoil band is indicated by a red star.      
     \label{leakageFig} }
\end{figure}

Current results from the CDMS experiment are based on 74.5 live days of data from six (250g each) Ge and six (100g each) Si detectors.  This amounted to 96.8 kg-days in germanium and 31.0 kg-days in silicon before rise-time criteria were applied.  The signal region was defined using neutrons from a $^{252}$Cf source.  The data quality and rise-time selection criteria were set and leakages and efficiencies were calculates without looking at the WIMP-search data in the signal region.  The number of  events estimated to pass the selection criteria  was 0.4 $\pm$ 0.2(stat) $\pm$ 0.2(syst) in Ge and 1.2 $\pm$ 0.6(stat) $\pm$ 0.2(syst) in Si.  The number of neutron events estimated from GEANT4 simulations of the detectors and shielding was 0.06 in Ge and 0.05 in Si.  After timing cuts six events remained of which only one event was inside the nuclear recoil region as shown in figure \ref{leakageFig}.

Further investigation of this event revealed it to have occurred during a time of known poor detector performance.  Hence, it is not evidence for a WIMP.   This event was included in the limit calculation.

The upper limit on the WIMP-nucleon spin-independent cross section from CDMS is $1.6 \times 10^{43}$ cm$^2$ for a WIMP with mass of 60 GeV/c$^{2}$.  This is a factor of ten lower than any other experiment.  As can be seen in figure \ref{limitsFig}, this limit excludes large regions of SUSY parameter space under some frameworks.

Currently CDMS is commissioning three additional detectors.  Improvements have also been made to the cryogenics, background rejection and data aquisition system.   In the next run CDMS plans to run a total of five towers each containing six detectors for a total of 30 detectors.   This brings the detector mass to  4.75 kg of Ge and 1.1 kg of Si.  The expected gain in sensitivity is a factor of 10 when run through the end of 2007.

In the future the collaboration plans to build a 25 kg experiment called SuperCDMS composed of cryogenic Ge detectors.  This experiment is proposed to be built in SNOlab at a depth of 6060 mwe, which will reduce the muon flux by a factor of 500 and the high-energy neutron flux by approximately a factor of 100.   Currently, SuperCDMS is in the process of fabricating  prototypes for a 3-inch diameter, 1-inch thick detectors.
\section{Edelweiss}
The Edelweiss experiment ls located in the Fr\'{e}jus Underground Laboratory under the French-Italian Alps at a depth of 4800 mwe.  Similar to CDMS, Edelweiss measures both phonons and ionization.  However, instead of using TESs to measure phonons, Edelwiess measures heat from phonon interactions using a neutron transmutation doped (NTD) Ge thermometric sensor.  The measured neutron flux in the Fr\'{e}jus lab is $1.6 \times 10^{-6}$ n/cm$^2$/s for neutrons with energy less than 1 MeV.

\begin{figure}[t]
  \vspace{9.0cm}
  \includegraphics{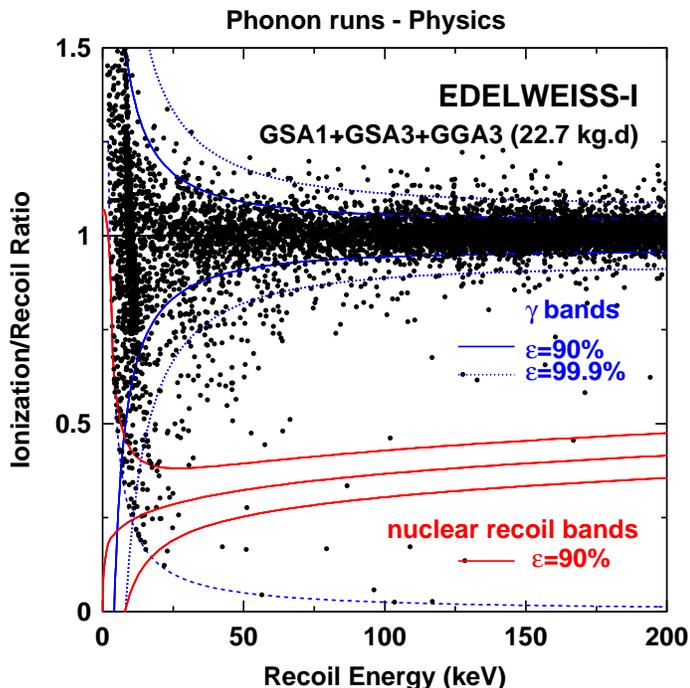}
  \caption{\it
   Distribution of the ionization/recoil energy ratio vs recoil energy  for the sum of the three detectors used in run 2003p\cite{edelweiss}.  Solid lines indicate the $\pm 1.65$ $\sigma$ electron and nuclear recoil bands.  Dotted lines indicated the $\pm 3.29$ $\sigma$ electron recoil band.  The ionization energy threshold is  shown by the hyperbolic dashed line.
    \label{edelweissFig} }
\end{figure}

The most current Edelweiss results are from three 320g Ge cryogenic detectors run in 2003 giving a fiducial exposure of  48.4 kg days and one 320g Ge cryogenic detector which ran from 2000-2003 giving a fiduial exposure of 13.6 kg days.   In the 62 kg days of fiducial exposure 40 nuclear recoil candidates were observed in the energy recoil region less than 15 keV and 3 nuclear recoil events were observed in the energy recoil region between 30 and 100 keV see figure \ref{edelweissFig}.   This gives Edelweiss a maximum sensitivity of $1.5 \times 10^6$ pb at 80 GeV.

Currently Edelweiss is in the process of building the second generation of their experiment, Edelweiss-II.  In 2004 they began installation of 21 Ge detectors of 320 g each which use the NTD heat sensor technology and 7 Ge detectors of 400 g each which will use a new NbSi thin-film sensor for detecting phonons.  They have also installed a new cryostat which is capable of holding 120 detectors.  Improvements made to the neutron shielding include adding a muon veto, adding 20 cm of lead and increasing the polyethylene shielding to 50 cm.
\section{CRESST-II}
The CRESST-II experiment is also located in Gran Sasso at a depth of 3500 mwe.   Similar to DAMA and KIMS, CRESST-II uses PMTs to measure scintillation light from particle interactions in their detector.  CRESST-II differs from these experiments in that they, like CDMS and Edelweiss,  also measure heat from phonon interactions.  The CRESST-II experimental apparatus consists of two 300g CaWO$_{4}$ crystals, two tungsten thermometers to measure heat from the phonon signals and PMTs to measure light.

The most current CRESST-II results are from a run lasting Jan 31 - Mar 23, 2004 which gave 20.5 kg days of exposure.  They observed 16 nuclear recoil events in total which is consistent with their background predictions\cite{cresst}.  It should be noted that this data was taken with no active neutron veto.
\begin{figure}[t]
  \vspace{9.0cm}
  \includegraphics{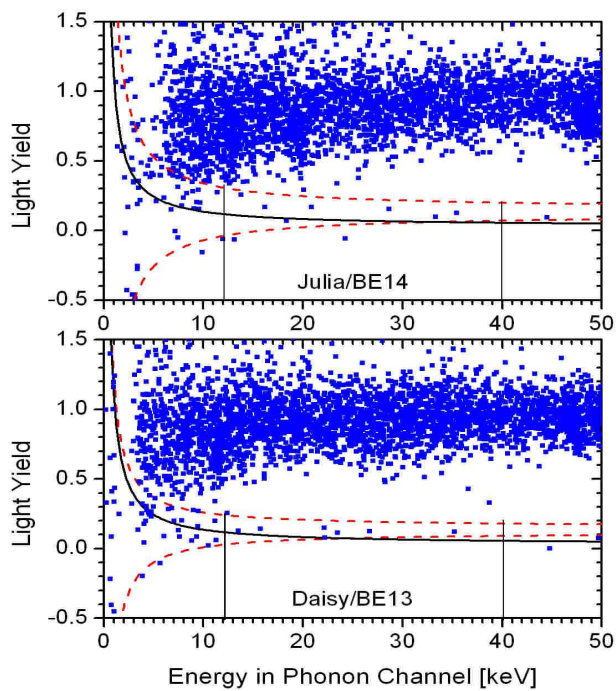}
 \caption{\it
   Light yield (scintillation/phonon energy) versus phonon energy for the two CRESST detector modules\cite{cresst}.  The region below the dashed curves contains 90\% of the nuclear recoils.  The solid vertical lines indicate the energy threshold for this analysis.
    \label{cresstFig} }
\end{figure}

In March of 2004 operation was stopped to install an active neutron veto and a 66 channel SQUID readout which will enable operation of 33 detector modules.  The work on the neutron veto and SQUID readout is now complete.  Current work on the detector-holder system and new analysis software is in progress.  The CRESST-II collaboration anticipates resumption of data-taking by the end of summer 2006.
\section{Zeplin}
The Zeplin-I detector is located in the Boulby Mine, United Kingdom at a depth of 2800 mwe.   The detector is composed of liquid Xe target with mass ~ 5 kg and photomultiplier tubes.  Differing from CDMS, Edelweiss and CRESST, Zeplin-I measures only one discrimination factor:  the pulse shape of the light.

The most recent Zeplin I results\cite{zeplin} are based on a total of 293 kg days of exposure.  The fiducial volume of the target mass is 3.2 kg.  The maximum sensitivity of the Zeplin I detector was $1.1 \times 10^{-42}$ cm$^{2}$.  No in situ calibration was preformed.

\begin{figure}[!t]
  \vspace{9.0cm}
  \includegraphics{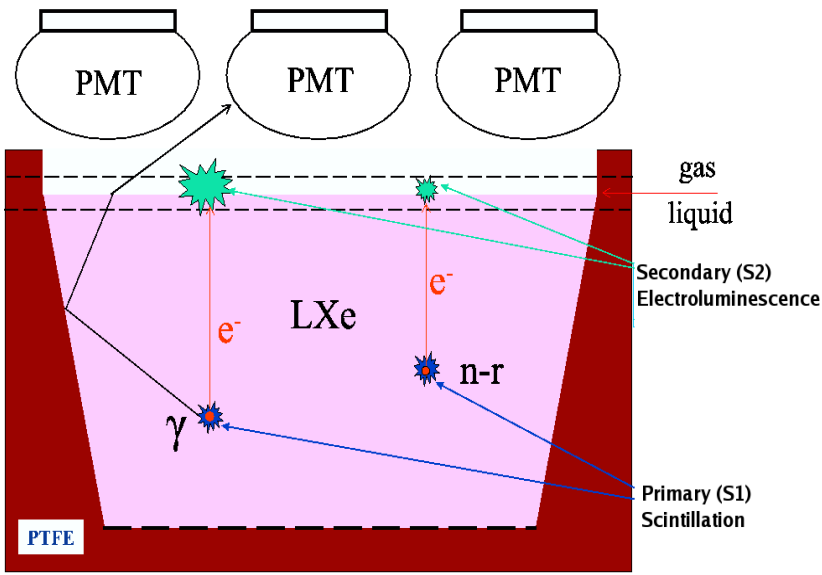}
  \caption{\it
  Schematic of Zeplin II\cite{zeplinII}.  Particles entering the tank will first interact in the liquid Xe producing a primary scintillation light signal.  The secondary electrons are then drifted by an electric field to the Xe gas layer.  Here the electrons give off a secondary electroluminescence.  Both of these signals are read out by seven PMTs of five-inch diameter located at the top of the detector.  The parent reaction can be identified as a nuclear or electron interaction by comparing the two signals.
    \label{zeplinIIFig} }
\end{figure}

The next phase of the Zeplin experiment, Zeplin II, is currently being commissioned in the Boulby mine.  Zeplin II differs from its predecessor in that it is a two-phase liquid-gas Xe detector with a fiducial volume of 30 kg.  A schematic of the detector can be seen in figure \ref{zeplinIIFig}.  The detector design is essentially a tank of liquid Xe with a layer Xe gas at the top with an electric field through the detector.  Photomultipliers on the top of the detector would record scintillation signals from particles interacting in the detector.   The detection principle is simply that a particle would interact in the liquid Xe giving off primary scintillation light.  The electric field would then drift secondary electrons into the gaseous region where they would produce a secondary electroluminescence.  The initial or parent particle would be identified by comparing the first and second scintillation signals.

Under parallel development is the Zeplin III detector.  It uses the same 2-phase idea as Zeplin II.  The detectors differ in that Zeplin III uses 31 smaller two-inch PMTs which are located at the bottom of the detector.  The total mass of the Zeplin III experiment is 8 kg.  The Zeplin III experiment is currently in the process of performing calibrations, system checks and optimizations.  They expect to run from 2007 - 2012 in the Boulby mine.

The Zeplin collaboration has started planning Zeplin IV/MAX which is envisioned to be a 1-ton experiment to be installed and run in SNOLab from 2008 - 2012.  

\section{XENON10}
The XENON10 experiment is another 2-phase Xe detector that is under commissioning in the Gran Sasso National Laboratory\cite{xenon}.  It has a total mass of 15 kg of liquid Xe.  XENON10 differs in design from the Zeplin detector family in that it has 48 eight-inch PMTs on the top and 41 eight-inch PMTs bottom of its detector.  The XENON experiment plans to begin running with shielding in 2006.
\section{Summary and Outlook}
\begin{figure}[t]
  \vspace{9.0cm}
  \includegraphics{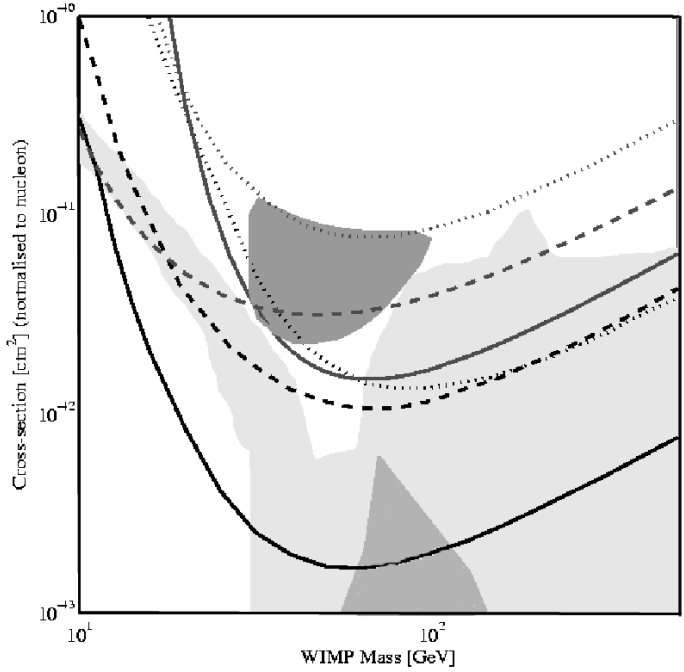}
  \caption{\it
    Current experimental results for the case of spin-independent WIMP-nucleon interactions.  DAMA is the only experiment to claim a signal (solid region).  The best limit is given by the CDMS-II experiment (solid black).  Limits from Zeplin I (dashed black), Edelweiss (dotted black), CRESST (solid grey), CDMS-II Si (dashed grey), and KIMS (dotted grey) are also shown.  Theoretical SUSY models taken from A. Bottino {\it et al} (light pink) and J. Ellis {\it et al} (light green).
    \label{limitsFig} }
\end{figure}

The field of direct detection dark matter physics is in a very active and exciting time.  Final results from many experiments world-wide including CDMS-II,  Edelweiss-I, CRESST-II and Zeplin-I have been released in the last year and are shown in figure \ref{limitsFig}.  DAMA claims to have observed a dark matter signal at 6.3 $\sigma$.  This claim has not yet been confirmed by another experiment.   All of these experiments are in the processes of commissioning the next stage of their detectors.  Several new detectors, including XENON10, are also coming online this year.

Plans for a ton-scale dark matter experiment are also being made.  Proposals are in progress by the CDMS, Zeplin and XENON collaborations.  Although not mentioned in this paper, the LHC will be coming online next year.  It too will search for dark matter and will be a great compliment to the ton-scale experiments.
\section{References}
\end{document}